\documentclass[aps,prl,onecolumn,superscriptaddress]{revtex4}

\usepackage[dvips]{graphicx}
\usepackage{latexsym}
\usepackage{graphicx}
\usepackage{epsfig}
\usepackage[english]{babel}
\usepackage[latin1] {inputenc}
\usepackage{amsmath,amsfonts,amssymb}
\usepackage{bm}

\newcommand \be {\begin{equation}}
\newcommand \bea {\begin{eqnarray}}
\newcommand \ee {\end{equation}}
\newcommand \eea {\end{eqnarray}}
\newcommand \bed {\begin{displaymath}}
\newcommand \eed {\end{displaymath}}

\newcommand{\bit}{\begin{itemize}}
\newcommand{\eit}{\end{itemize}}

\newcommand{\bgar}{\begin{eqnarray}}
\newcommand{\enar}{\end{eqnarray}}

\begin{document}

\title{ Growth Laws in Cancer: Implications for Radiotherapy}

\author{P.Castorina$^{(a,b)}$, T.S.Deisboeck$^{(c)}$,P. Gabriele$^{(d)}$ C.Guiot$^{(e,f)}$}
\affiliation{$^{(a)}$ Dipartimento di Fisica, Universita' di
Catania,
Italy\\  $^{(b)}$ INFN-Catania, Italy  \\
$^{(c)} $ Complex Biosystems Modeling Laboratory, Harvard-MIT (HST) Athinoula A. Martinos Center for Biomedical Imaging, 
Massachusetts General Hospital, Charlestown, MA 02129.
\\ $^{(d)} $ Institute for Cancer Research and Cure, IRCC, Candiolo (TO), Italy 
 \\ $^{(e)} $ INFM- Torino UNiversita' e Politecnico, Italy
\\ $^{(f)} $ Dipartimento Neuroscienze, Universita' di Torino, Italy}

\maketitle


\section{ABSTRACT}

Comparing both, the more conventional Gompertz tumor growth law (GL) and the ``Universal'' law (UL), recently 
proposed and applied to cancer,
we have investigated the growth law's implications on various radiotherapy regimen.
According to GL, 
the surviving tumor cell fraction could be reduced 'ad libidum', independently of the initial tumor mass,
 simply by increasing  the number of treatments. 
On the contrary, if tumor growth dynamics would indeed follow the Universal scaling 
law,  there is a lower limit of the survival fraction 
that cannot be reduced any further regardless of the total number of treatments. 
This finding can explain the so called ``tumor size effect'' and  
re-emphasizes the importance of early diagnosis as it implies that radiotherapy 
may be successful provided the tumor mass at treatment onset is rather small. Taken together with our previous works, 
implications of these findings include revisiting standard radiotherapy regimen and overall treatment protocols.

\vfill
\eject

\section{ INTRODUCTION}

A more detailed understanding of tumor growth is crucial for the clinical management of the disease and 
tumor size is a main determinant of clinical severity and a major factor of the staging criteria before and during 
radiotherapy (RT) \cite{greene}
Tumor regrowth during radiotherapy   is therefore an important clinical parameter \cite{kim}and, in particular,   
the dose-response relationship 
and thus the probability of treatment benefit critically depend on the tumor re-growth pattern 
in the interval between the fractional irradiation treatments.

To clearly evaluate  the clinical results ,the tumor cell 
'survival fraction' $S$ ,  after $n$ irradiations at dose per fraction $d$,  in the
overall treatment time $t$, is usually written as
\begin{equation}
-ln(S) = n (\alpha d + \beta d^2) - \gamma t
\end{equation}
and depends on  the tumor radiosensitivity,  
expressed  by the parameters  $\alpha$  and  $\beta$, , according to the linear-quadratic model,and on
the regrowth parameter  $\gamma = ln 2/ \tau_{eff}$, where $\tau_{eff}$  is the
 the average clonogenic doubling time \cite{Fowler}.
The above equation is, up to now, the basis for RT
scheduling, and would predict  the probability
$P$ of tumor control, defined as
$P= \exp(-c S)$, being $c$ the clonogen number. 

Untreated tumor growth has been usually described by means of the Gompertz law (GL)
\cite{steel,weldom,norton1,yorke}, a non linear growth pattern proposed a long time
ago in actuarial mathematics \cite{gompertz}.
Moreover, in a transplantable rat tumor, it was shown that control and regrowth curves after 
radiotherapy could be fitted by the same gompertzian law, 
provided adjustments for the initial lag and the estimated number of clonogens 
immediately after irradiation  were performed \cite{jungh}. Gompertzian growh has been assumed 
to describe human tumor repopulation during fractional radiotherapy also by Hansen et al. \cite{hanseno} and 
by O'Donougue \cite{O'Donogue}.

 Recently, an alternative general growth law, based on the scaling properties
of the nutrient supplying distributive network has been proposed \cite{west1,west2} which is
claimed to be ``Universal'' since it is able to fit most living organisms' growth pattern, covering more 
than 27 orders of magnitude in mass. Since then, their Universal law (UL) has been shown to fit reasonably
well many available data on tumors in vivo and on multicellular tumor spheroids (MTS) \cite{guiot1}.

In this paper we consider a close analysis of the two different growth patterns aiming at evaluating 
their impact on clinical treatment regimen. Our results, clinically useful ``per se''', permit to understand
some observed, but still unclear, effects. 
 
\section{ TUMOR GROWTH LAWS}

Up to 1956 \cite{collins} human tumor growth was simply
described as ``slow'' and ``rapid'' , without any attempt
for quantitative description \cite{retsky}. A naive view would
consider an exponential growth, from a 10 microns cell to a 1 liter
tumor in about 20 doublings. On this basis, from two measurements of volume $V_1$ and $V_2$ at
different times $t_1$ and $t_2$, the constant tumor doubling time
 can be estimated as:  $\tau_d  = (t_1 -  t_2) / ln 2 (V_1 / V_2)$.
Several studies on animal models \cite{steel} and  a couple of
very important
 investigations on breast and prostate cancer in humans \cite{norton1,yorke}
 showed that,  far from
being constant, $\tau_d$ was seen to change
during the tumor growth,  which is mathematically  well described  by a Gompertzian
growth kinetics   \cite{gompertz,steel,weldom}
\be
N(t) = N_o^g \exp{[\frac{a_o}{K_g} (1 - exp(-K_g  t ) ]}
\ee 
where $N(t)$ is the number of cells, that is proportional to the tumor mass,
$K_g$ and $a_o$ are constants and  $N_o^g = N(0)$.

Although it  is generally considered as a phenomenological tool, there are many attempts to derive the
Gompertz law  by   more fundamental dynamics \cite{zeliko,io}. In the analysis of in vivo tumor growth a single set of
growth parameters is insufficient to describe the clinical data. Tumor cells have different growth conditions and 
characteristics in different patients and the variation of tumor growth in patient population  has been modeled by
using a distribution of growth parameters. It turns out that the data are fitted  by a log-normal distribution 
of the parameter $K_g$. For example the Bloom data on breast cancer \cite{bloom} are consistent with $N_o^g = 4.8 * 10^{9}$,
 $N_\infty^g = 3.1 * 10^{12}$, a mean value of the log-normal 
distribution given by $ln(K_g) =  -2.9$ and a standard deviation  $ln(K_g) =   0.71$  \cite{norton1}.

A new model of the tumor growth has recently been proposed on the basis of the paper by
West et al. \cite{west1} that, regardless of the different masses and development times,
shows that many living organisms share a common growth pattern and, provided masses and growth times are properly
rescaled, the same universal exponential curve fits
their ontogenic growth data. 
This phenomenon is explained 
by basic cellular mechanism \cite{west2} assuming a common fractal pattern in the vascularization of the investigated taxa.
 More precisely, starting from a cell number $N_0^w$ ( or mass $M_0$) at birth,
$N$ (or $M$) increases, with decreasing rate, up to a maximum value
$N_\infty^w$ (or $M_\infty$). Introducing the ratio
$r=(N/N_\infty)^{\frac{1}{4}}= (M/M_\infty)^{\frac{1}{4}}$, 
i.e. the relative proportion of total energy expenditure required to ensure  maintaince,
the general growth pattern, that we call Universal Law (UL), follows 

\begin{equation}
r = 1 -  exp (- \tau_W),
\end{equation}
where
\be
\tau_W= \frac{\sigma t}{4  {M_\infty^w}^{1/4}} - \ln (1 - r_0),
\ee 
 $r_0=(M_0/M_\infty^w)^{\frac{1}{4}}$ and $\sigma$ is a constant fitted by data, , with dimension $g^{1/4}$/month
when $M$ is the tumor mass in $g$.

 Guiot et al. \cite{guiot1} applied this growth pattern to tumors, satisfactorly
fitting  MTS data, as well as for experimental rats and mouse tumors and finally
for human breast and prostate cancer.
Contrary to GL, the UL has never been applied to the case of irradiated tumors. 

According to the standard clinical procedure,
 the treatment dose $d$ is given at regular intervals. Let us assume that the
surviving fraction for clonogenic cells is
given by the linear-quadratic model ( see eq.1) and
the repopulation specific rate, $\lambda$ of the clonogenic cells is a
function of the population size $\lambda(N(t))$.

Therefore the differential equation for the considered irradiated system is 
\begin{equation}
\frac{1}{N}\frac{dN}{dt}=  \lambda (N(t)) - \Sigma_{j=1}^{n-1} (\alpha d +
\beta d^2) \delta ( t - j \tau)
\end{equation}
where $\tau$ is the interval between two treatments,
 $n$ is the number of treatments fraction given by time $t \geq  (n-1) \tau$.

For an exponential growth, i.e. constant rate $\lambda(N(t))= \gamma =
\frac{\ln 2}{T}$, one obtains eq. (1) with
$\tau_{eff}= T$, the doubling time of the exponential law.

For both Gompertz and Universal laws a more detailed analysis ( see Appendix) is needed
to evaluate  the difference between  the two 
growth patterns  in the  survival fraction $S$ after a realistic irradiation
treatment.

In addition to standard treatment ( up to $70$ Gy with daily doses of $2$ Gy) we investigated also non-standard 
treatments schedules recently proposed in the clinical literature. In particular we considered 
the so called 'hyperfractionation', consisting in 3 daily doses of $0.8$ Gy for a total of  $60$ Gy 
in 4 weeks, 'hypofractionation' (5 Gy x 5 days for a total dose of 25 Gy in 1 week) and 'CHART' protocol 
(1.5 Gy three times a day for a total dose of 54 Gy).

\section{ Results}

After the initial phase, the ln(S) computed according
the UL can be  reduced only by changing applied dose and interval, yet cannot be further reduced 
by increasing the number of treatments.
This is a strong difference with respect to Gompertz growth where  
the final survival fraction can  be always reduced 
by increasing the number of treatments.

As an example, Fig. 1 shows the $ln(S)$ vs. the number of treatments when
$d=2$ $Gy$, $\tau=1$ day (interval between two treatments),
 $\alpha=0.3$ $Gy^{-1}$ and $\alpha/\beta=10$ $Gy$ (breast cancer).
GL prediction does not depend on the actual tumor mass, 
while UL prediction does. Tumor ( asymptotic) final mass M is assumed $\simeq 640$ g \cite{norton1}. 
Since M is a parameter of the West law, it is convenient to define the tumor mass as a fraction of M. 
Two cases are considered:  the empty romboids refers to a very small tumor
whose mass is $1\%$ of the final one and the empty triangle to a small tumor, whose 
mass is $10\%$ of the final one.  

It is apparent from the
figure that, according to GL, the surviving tumor cell fraction could be reduced 'ad libitum', simply
by increasing the number of radio-therapeutic fractionated treatments, independently of the initial
tumor mass \cite{greene,O'Donogue}.On the contrary, the UL establishes a lower limit for the
survival fraction  that cannot be reduced any further
regardless of the total number of treatments.

In particular, while in the first half of the treatment only a small discrepancy is observed, 
 approaching the final standard total dose of $70$ $Gy$ ( or 35 treatments) 
 the predicted values for $ln(S)$ 
by the UL law is almost 7 order of magnitudes larger than expected by the GL.
In other words, therapeutic control of tumor proliferation is poorer if 
cellular regrowth follows UL instead of GL unless the total dose needed for eradication is small enough 
to be in the range where Gl and UL predict the same value for S. Since such small doses are required 
only for very small tumors, the  UL may be able to explain the so called ``tumor size effect'', i.e. 
why the tumor control rate achieved by radiation treatments alone rapidly declines for large tumors 
( T3 or T4 or N2c/N3 in the clinical practice).

\begin{figure}
\epsfig{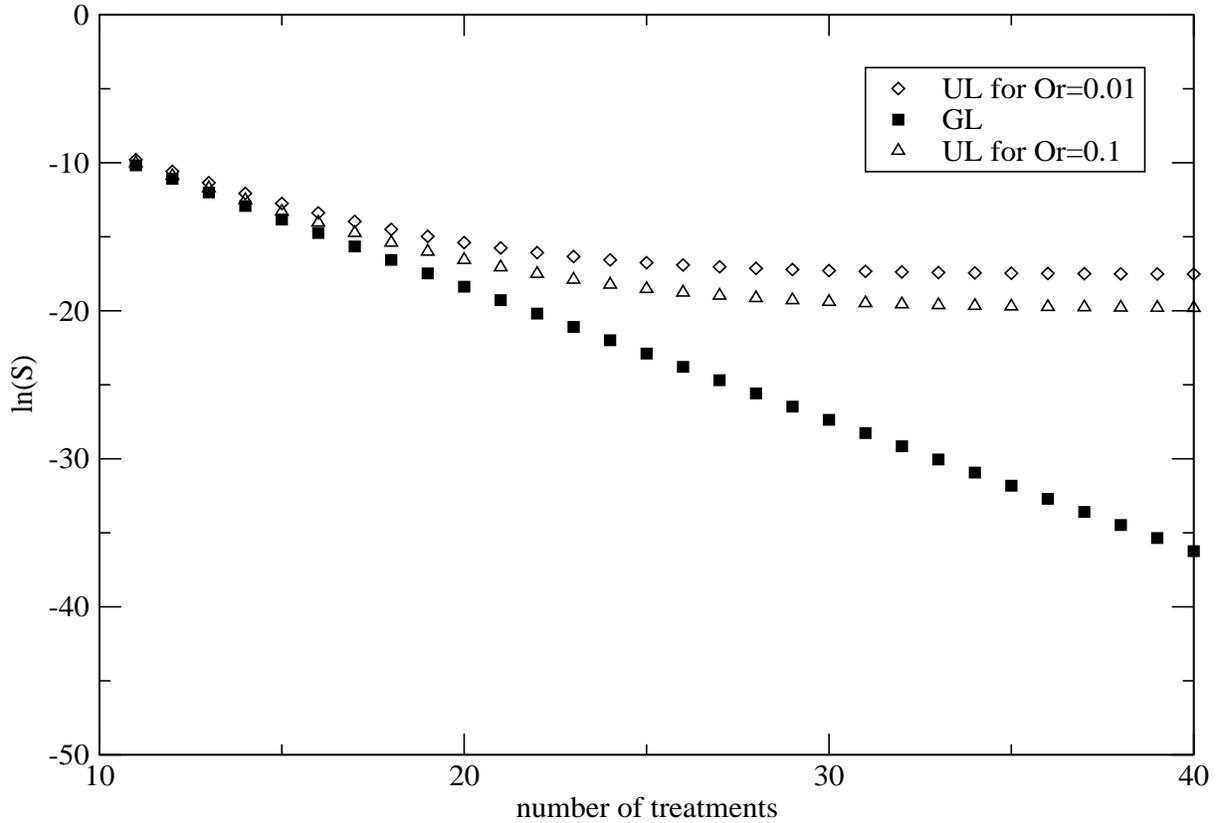}
\caption{$\ln(S)$ vs  the number of  treatments when $d=2.0$ Gy
, $t=1.0$ day and different tumor initial mass,
expressed as the percentage , $Or$, of the final tumor mass for breast cancer}
\end{figure}

In order to stress the different impact of GL and UL in the case of standard treatment for tumors of different
volume, we computed P at different number c of clonogens: $10^3$, $10^5$ and $10^7$ respectively( Fig. 2)

\begin{figure}
\epsfig{file=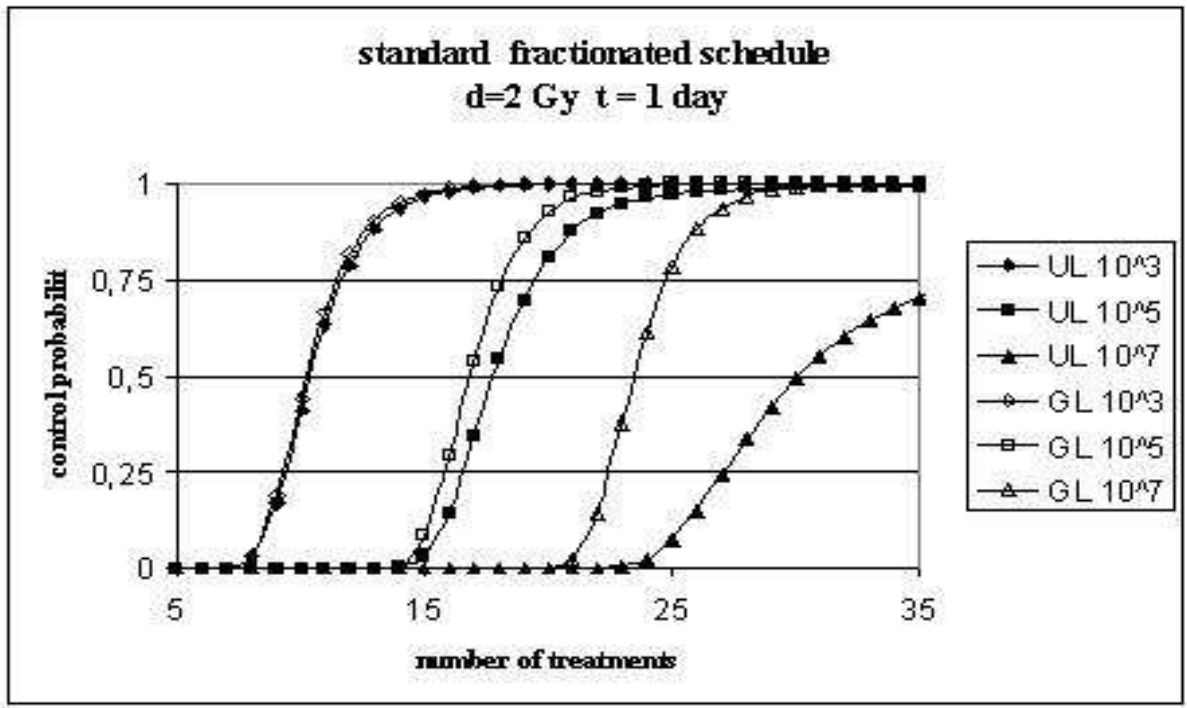,height= 12.0 true cm,width=16.0 true cm, angle=0}
\caption{$P$ vs  the number of  treatments when $d=2.0$ Gy
, $t=1.0$ day and different tumor clonogens number c}
\end{figure}

As expected, while at low clonogen number c both growth laws predict the same control probability, at intermediate 
c the therapy success is delayed and at large c is unattained .

The previous interesting clinical results  are further investigated by considering  the UL and the GL
with different treatment schedules.
In particular,since  clinical experience confirms that highly proliferative tumors are unsatisfactorily 
treated by conventional RT schedule,
 we have performed simulations by assuming non-conventional, yet widely applied RT schedules, such as 
 hyperfractionation 
\cite{acchyper}, CHART protocol \cite{chart} and hypofractioned regimes \cite{hypo}, 
which are known to be more effective in controlling the evolution of highly proliferative tumors.

In figs 3, 5 and 7 are  reported the  values of the final survival fraction ( in log scale) 
as a function of the number of treatments   for hypofractionation,  hyperfractionation 
and  CHART protocol   by considering the
regrowth according the GL and the UL, when the initial observed tumor mass  is respectively 1 and 10    percent of the asymptotic value 
(i.e. the maximum size attainable from this specific tumor), i.e.  $Or=N_{in}/N_\infty= 0.01,0.1    $.
The same shedules have been investigated for cure probability assuming the tumor mass to
 be $10\%$ of the final one and clonogenic number equal to $10^3$, $10^5$ and $10^7$. 
Figs 4, 6 and 8 are referred to hypofractionation, hyperfractionation and CHART respectively.

\begin{figure}
\epsfig{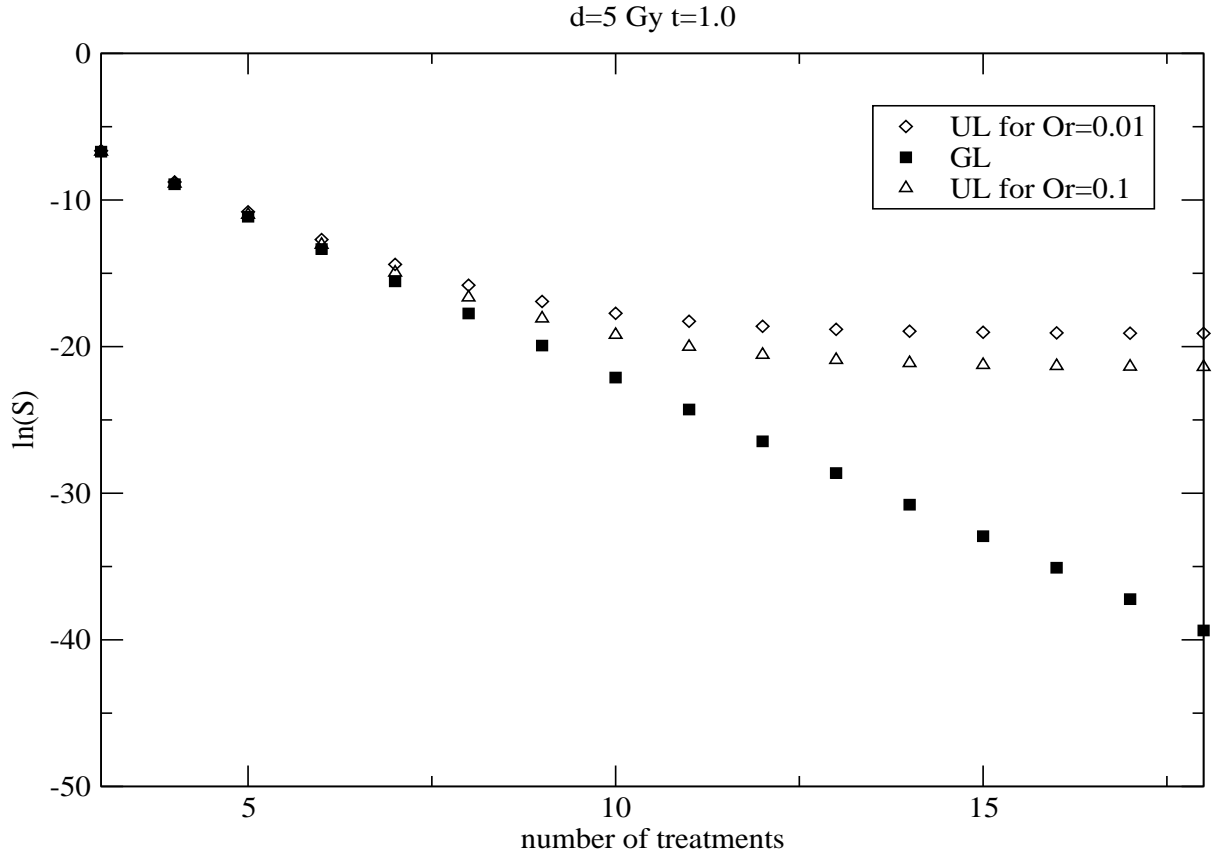}
\caption{$\ln(S)$ vs  the number of  treatments when $d=5$ Gy
, $t=1$ day and  different initial tumor mass,
expressed as the percentage of the final tumor mass , $Or$, for breast cancer}
\end{figure}

\begin{figure}
\epsfig{file=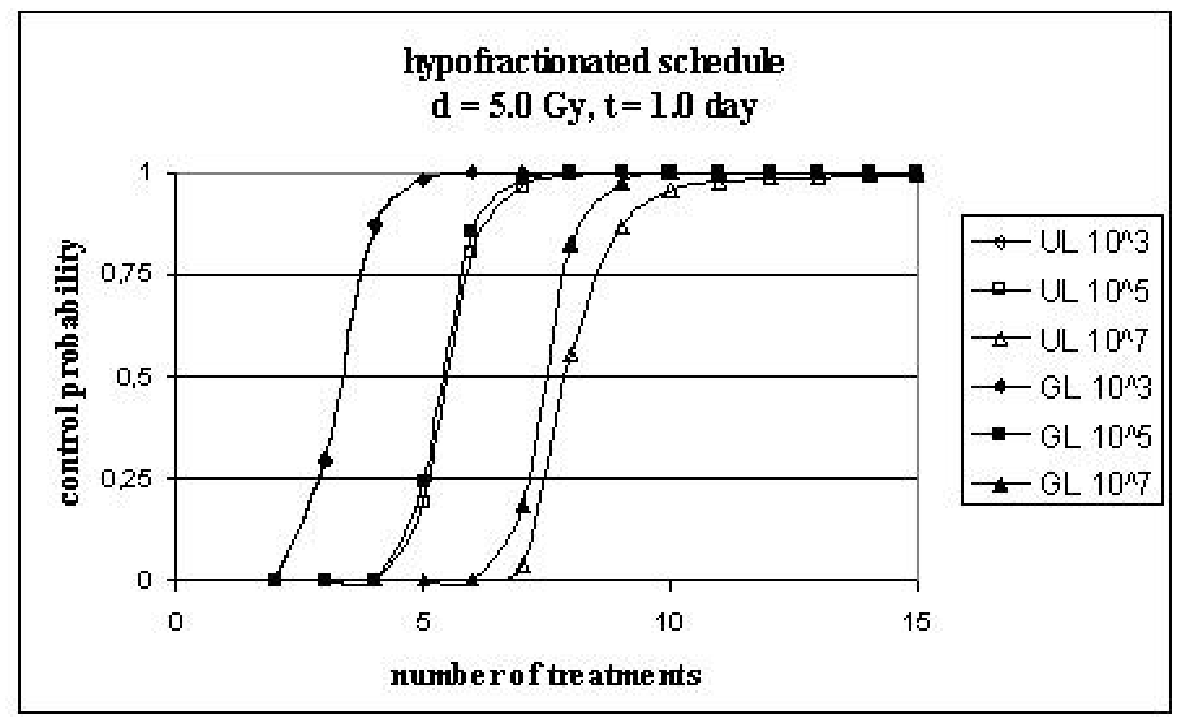,height= 12.0 true cm,width=16.0 true cm, angle=0}
\caption{$P$ vs  the number of  treatments when $d=5$ Gy
, $t=1.0$ day and different tumor clonogens number c}
\end{figure}

\begin{figure}
\epsfig{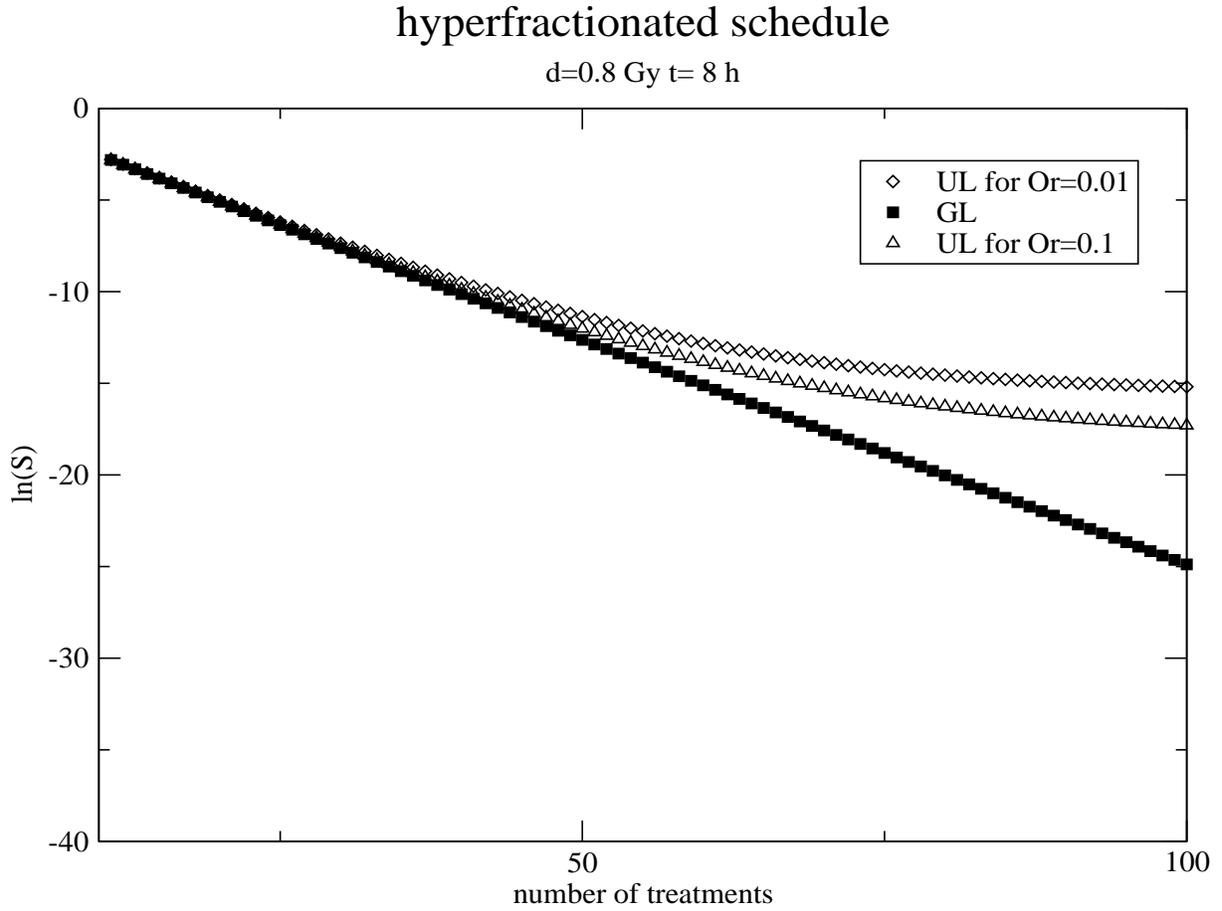}
\caption{$\ln(S)$ vs  the number of  treatments when $d=0.8$ Gy
, $t=8$ h. and different initial tumor mass,
expressed as the percentage of the final tumor mass , $Or$, for breast cancer}
\end{figure}

\begin{figure}
\epsfig{file=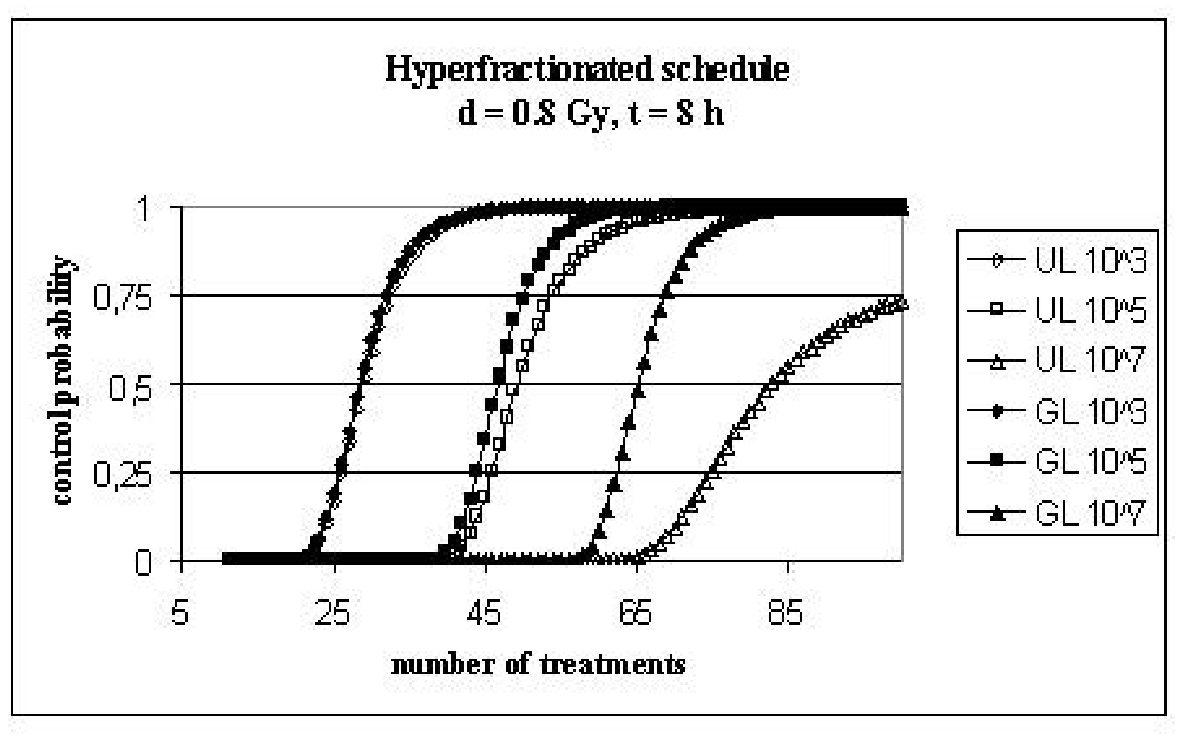,height= 12.0 true cm,width=16.0 true cm, angle=0}
\caption{$P$ vs  the number of  treatments when $d=0.8$ Gy
, $t=8$ h. and different tumor clonogens number c}
\end{figure}

\begin{figure}
\epsfig{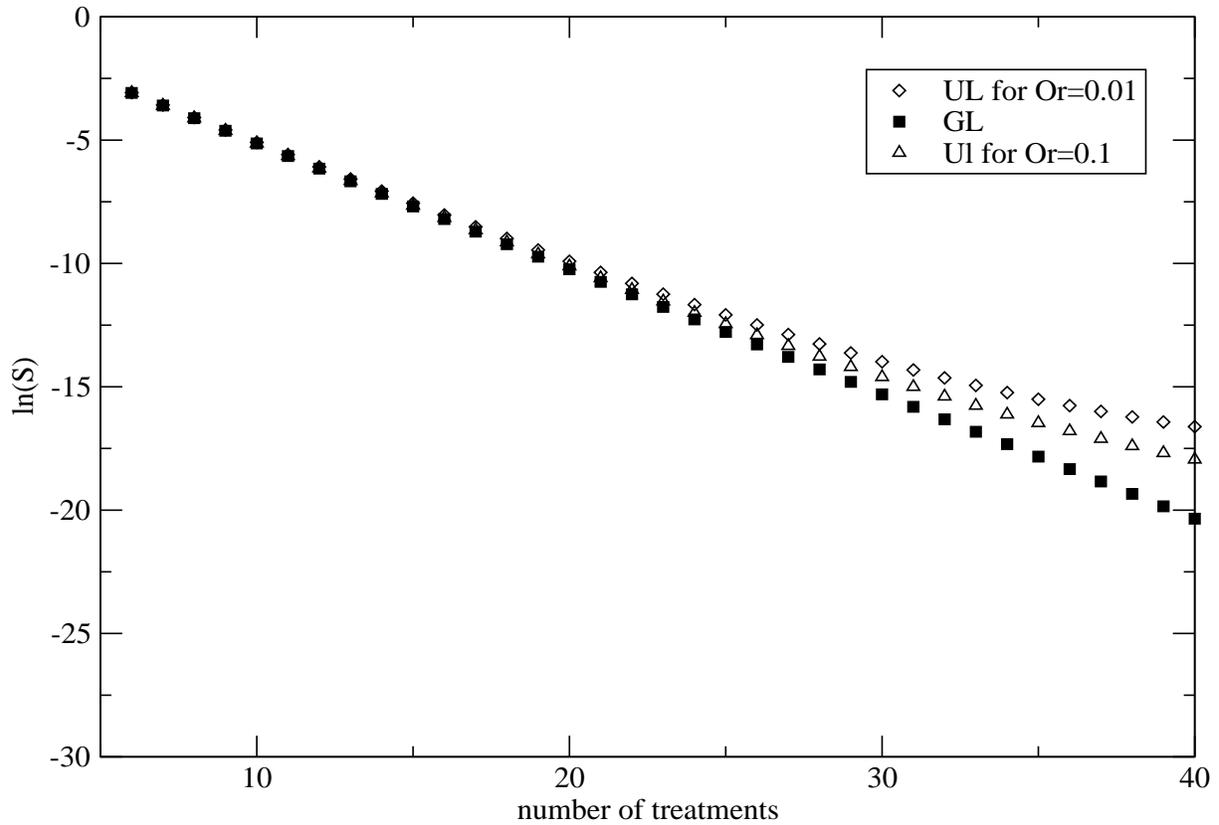}
\caption{$\ln(S)$ vs  the number of  treatments when $d=1.5$ Gy
, $t=8$ h.  and different initial tumor mass,
expressed as the percentage of the final tumor mass , $Or$, for breast cancer}
\end{figure}

\begin{figure}
\epsfig{file=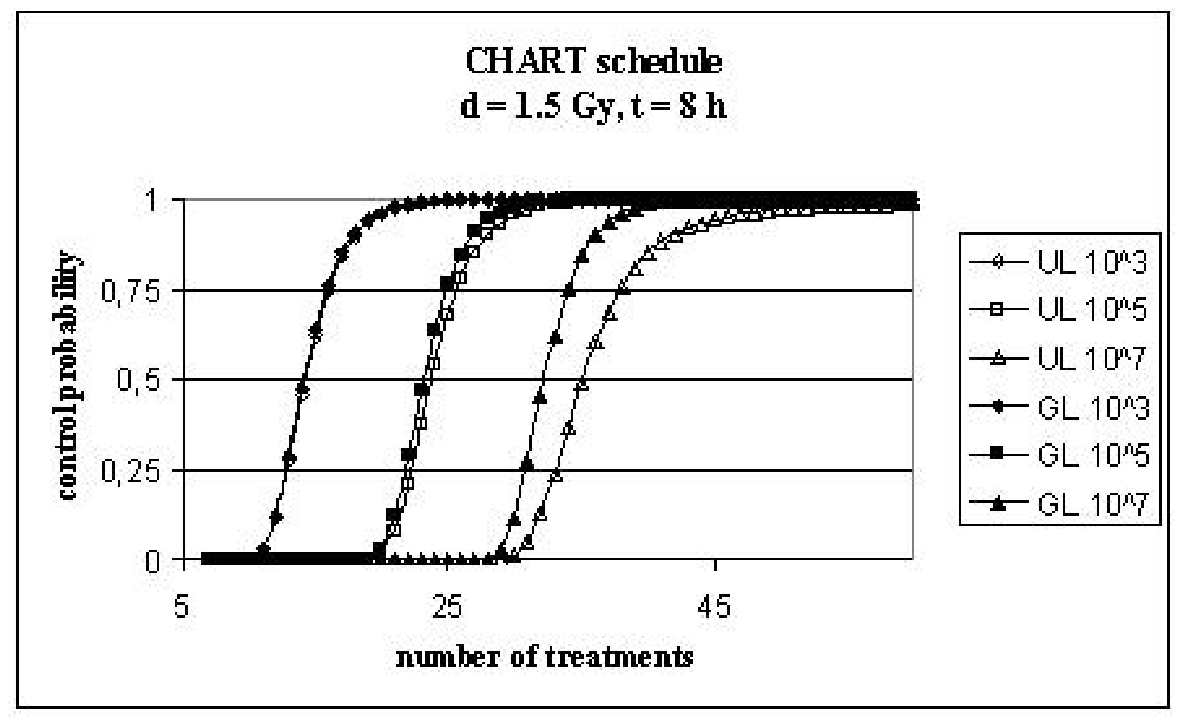,height= 12.0 true cm,width=16.0 true cm, angle=0}
\caption{$P$ vs  the number of  treatments when $d=1.5$ Gy
, $t=8$ h. and different tumor clonogens number c}
\end{figure}

For Hypofractionation and CHART shedules 
almost the same results are obtained for s
mall and intermediate c, while complete
 therapeutic success can be achieved for tumors following the UL provided a larger number of treatments is delivered. 
In the case on hyperfractionation, 
on the contrary, tumor following the UL cannot be 
satisfactorily treated it c is large, and there is no advantage with respect to the standard schedule.

\section{ Discussion and Conclusions}

In this paper we compare the tumor survival fraction during RT regimen predicted by 
the Gomperz Law (GL) and the Universal  growth law (UL) , based on scaling principles.
We note that the survival fraction, $S$,critically depends on the
tumor re-growth rate. According to GL, the surviving tumor cell fraction could be reduced 'ad libitum', simply
by increasing the number of radio-therapeutic fractionated treatments, independently of the initial
tumor mass \cite{greene,O'Donogue}.On the contrary, the UL establishes a lower limit for the
survival fraction, weakly dependent on the clonogenic number, that cannot be reduced any further
regardless of the total number of treatments.

Two important considerations follows:

1) Predictions of tumor regrowth by GL and UL are very similar only in the initial part of the
treatment, i.e. up to around 25-30 Gy. 
In other words, only if the tumor mass is small enough to be cured by an overall dose delivered by few
treatments, both the GL and the UL predict similar outcomes. 

Provided the number of clonogenic cells is accordingly small,
a decrease of about 10 units in the ln(S) already eradicates the tumor and RT reaches its goal 
independently on the actual re-growth curve followed by the tumor. Results are no more satisfactory 
when larger tumors are irradiated, because, contrary to expectations relying on the GL, ln(S) doesn't 
decrease any more, and clonogenic cells are not definitively eliminated.

 The  ``tumor size effect'' can be therefore understood on the basis of the UL.
The dependence of the surviving fraction on the tumor volume was already observed by Stanley et al in 1977 
in lung tumors \cite{stanley}, 
and re-emphasized by Bentzen et al and Huchet et al in \cite{bentzen,huchet}.
Larger tumors are expected to have a higher number of clonogenic 
cells to be killed 
as well as a more hypoxic environment. Both factors affect  tumor regrowth as well as, possibly, 
tumor radiosensitivity.
While GL is volume-insensitive, UL accounts for the tumor growth stage, predicting different survival
fractions after RT treatments.

Moreover, as far as the local control is concerned, this is qualitatively in  agreement with the results  
for instance in breast cancers treated by radiotherapy alone where the only two significant factors determining treatment
outcome (control vs. failure) are the overall dose and the tumor size \cite{arri}.
Furthermore, a more recent analysis on breast cancer \cite{ver} shows that, even in the presence
of nodal involvement, tumor size does not lose its prognostic role, rather it maintains its predominant 
effect on mortality. It is noteworthy
that in many pathologies the dose for $90\%$ local control is strictly related to tumor volume:
 for instances, in human malignant
epithelial tumors, it ranges from $50 Gy$ for small lesions to $60 Gy$ for linear dimensions 
$< 2 cm$ to $75 Gy$ for large lesions 
($4-6$ cm in min-max diameter). Finally, rapid tumor re-growth during ``long'' $(5-8 wk)$ 
radio-therapeutic treatment is an
 important clinical parameter \cite{wit}
This fact re-emphasizes the importance of early shrinkage of the gross tumor mass, i.e. 
by surgical debulking prior to
radiation treatment, since it implies that radiotherapy may be successful provided the tumor mass at
treatment onset is rather small.

2) When larger tumors are considered, we would expect that, according to GL, therapeutic results
 depend on the total delivered dose, independently on the actual schedule. Tumor regrowth according to the UL, on
the contary, shows a dependence on different RT schedules.

Actually, clinical experience confirms that highly proliferative tumors are unsatisfactorily treated by 
conventional RT schedule.
Simulations are therefore proposed by  assuming non-conventional RT schedules such as accelerated hyperfractionation 
\cite{acchyper}, CHART protocol \cite{chart} and hypofractioned regimes \cite{hypo}. 

 Our simulation shows that for tumors following the UL law there is a therapeutic advantage is using Hypofractionation and 
CHART schedules, since a complete success can be achieved even for large and/or very aggressive tumors ( c large),
while hyperfractionation doesn't improve results with respect to standard RT schedule.

As far as Hypofractionation and CHART are concerned, a good agreement between the model and the clinical results is found,
since both schedules are satisfactorily used in palliation and in treating advanced neoplasies.
Regarding hyper-fractionation, attention should be paid to the treatment details. The delivery of 0.8 Gy three time a day 
('plain' hyperfractionation) is actually performed with significant improvements in local control and survival probability in 
medium-size oropharingeal tumors \cite{orofar}.
Larger tumors are treated using a variety of schedules, such as the 'accelerated hyperfractionation ( 1.5-1.6 Gy twice a day) 
( there is evidence that for some tumors (inflammatory breast cancer \cite{infla},
head and neck cancer \cite{neck}) standard RT treatment may be accelerated with benefit) ,or using the so-called 
'concomitant boost' (by adding 1.2 Gy 
each day in the second and fifht weeks of treatment). 
The main concern in increasing the radiation dose is its impact on healthy tissue which should be 
spared  as
much as possible. The goal however can nowadays be achieved by 3D conformal radiotherapy 
in all its techniques (3DCRT, IMRT, Stereotactic treatment) allowing larger doses to be used. 
Recently these two options (accelerated fractionation and IMRT) are been used together in a 
particular in the 
SMART (simultaneous modulated accelerated radiation therapy) \cite{smart} or 
SIB (Simultaneous Integrated boost) \cite{sib}. Investigating all the above options is quite demanding, so they will
be targeted in a following paper.

 Since there is clinical evidence for better responses to some non-conventional schedules of large tumors, such as hypofractionation 
and CHART, the UL model may be more appropriate to account for
 tumor regrowth of highly proliferating  tumors during RT, 
and may help to logically explain clinical results.

Up to now, also the aforementioned RT regimens have not yet been investigated exhaustively with theoretical models
 and, to our knowledge a comparison between different growing tumors and/or different RT schedules is still missing.
 We think that, as for the tumor size effect,  the Universal Law can help in understanding the experimental data not explained by
 the Gompertz law. 

ACKNOWLEDGEMENTS: This work has been supported in part by NIH grants CA 085139 and CA 113004 and by the Harvard-MIT 
(HST) Athinoula A. Martinos Center for Biomedical Imaging and the Department of Radiology at Massachusetts General Hospital.

\section{Appendix}

Let us consider that an 'in vivo' tumor, with $N_{in}$ initial cells, is irradiated
at $t=0$  with a dose $d$ which istantaneously produces a survival fraction $S_0$, i.e.
\be  
N(0)=N_{in} \exp{[-(\alpha d+ \beta d^2)]}= N_{in} S_0
\ee

One can easily shows that, after $n$ equal treatment, the final survival fraction, $S_g= N(t)/N_{in}$,
 for the Gompertz  pattern  turns out to be:
\be
S_g = \exp{[-n(\alpha d+ \beta d^2) + R_g G]}
\ee
where  
\be
R_g = 1 - \exp{(-K_g \tau)},
\ee
\be
G = \Sigma _{i=1}^{m} (1-R_g)^{m-i} \ln{\frac{N_\infty}{N_{in} (S_0)^i}},
\ee
with  $m=n-1$.

For the West law the result is
\be
S_w = \exp{[-n(\alpha d+ \beta d^2)]} [R_w^{m} W]^4
\ee
where
\be
R_w=\exp{(-\frac{\sigma \tau}{4  {N_\infty^w}^{1/4}})},
\ee
\be
W= 1 + \frac{(1-R_w)}{(N_i/N_\infty)^{1/4}}\frac{1-1/\delta^m}{\delta - 1}
\ee
and
$\delta = R_w S_0^{1/4}$.

The costant $\sigma$ is small ( $\sigma \simeq 0.42$) $g^{1/4}$/month  and $\tau$ is typically between 1-2 days. Therefore,
for a typical dose of $d=2.5$ Gy,  $\alpha=0.3$ and $\alpha/\beta \simeq 10$ (breast cancer),  to the first order approximation,
 one can write $R_w= 1 -\epsilon$ with $\epsilon=\gamma \tau /4  {N_\infty^w}^{1/4}$ and  $R_w S_0^{1/4} \simeq S_0^{1/4} \simeq 0.4$.
 For a large number of treatments
( formally for $m \rightarrow \infty$), it finally turns out
 \be
S_w^{asy} = \frac{S_0}{ N_{in}} (\frac{\sigma \tau}{4(1 - S_0^{1/4})})^4 
\ee
 
Since eqs.(10-12) rapidly saturates to its asymptotic value,  there is a theoretical limit to  tumor control due to regrowth
according to  the UL which, by assuming for sake of simplicity $N_{in}=c$, is 
given by
\be
P_{asy}= \exp{(- S_w^{asy} N_{in})}= \exp{[-(S_0 (\frac{\gamma \tau}{4(1 - S_0^{1/4})})^4)] }
\ee

which is independ on  $N_{in}$, on $N_{\infty}$ and depends only on the dose, on the interval between treatments
and on the growth rate parameter $\sigma$. $P_{asy}$
 cannot be further improved by increasing the number of treatments but only changing the dose and the scheduled interval.
This is a strong difference with respect to Gompertz growth where one can (in principle, always) reduce the final survival fraction 
by increasing the number of treatments. Indeed,
according 
to GL the
  final survival fraction is practically independent on $N_{in}$ and can be continuously decreased by increasing the number of treatments.
For the UL   there is a dependence on the initial cell number but there is no way, at fixed $d$ and $\tau$,
 to decrease  $P_{asy}$                      beyond its asymptotic value.

\end{document}